# ALL-MASK: <u>A</u> Reconfigurable <u>L</u>ogic <u>L</u>ocking Method for <u>M</u>ulticore <u>A</u>rchitecture with <u>S</u>equential-Instruction-Oriented <u>K</u>ey

Jianfeng Wang, *Student Member, IEEE*, Zhonghao Chen, *Student Member, IEEE*, Jiahao Zhang, *Student Member, IEEE*, Yixin Xu, *Student Member, IEEE, Tongguang Yu, Student Member, IEEE*, Enze Ye, *Student Member, IEEE*, Ziheng Zheng, Huazhong Yang, *Fellow, IEEE*, Sumitha George, Yongpan Liu, *Member, IEEE*, Vijaykrishnan Narayanan, *Fellow, IEEE*, and Xueqing Li, *Senior Member, IEEE*

*Abstract*—Intellectual property (IP) piracy has become a non-negligible problem as the integrated circuit (IC) production supply chain is becoming increasingly globalized and separated that enables attacks by potentially untrusted attackers. Logic locking is a widely adopted method to lock the circuit module with a key and prevent hackers from cracking it. The key is the critical aspect of logic locking, but the existing works have overlooked three possible challenges of the key: safety of key storage, easy key-attempt from interface and key-related overheads, bringing the further challenges of low error rate and small state space. In this work, the key is dynamically generated by utilizing the huge space of a CPU core, and the unlocking is performed implicitly through the interconnection inside the chip. A novel low-cost logic reconfigurable gate is together proposed with ferroelectric FET (FeFET) to mitigate the reverse engineering and removal attack. Compared to the common logic locking methods, our proposed approach is 19,945 times more time consuming to traverse all the possible combinations in only 9-bit-key condition. Furthermore, our technique let key length increases this complexity exponentially and ensure the logic obfuscation effect.

*Index Terms*—Hardware security, IP piracy, logic locking, key, ferroelectric FET (FeFET), reconfigurable circuit.

## I. INTRODUCTION

Integrated circuit supply chain has become worldwide since the designers tend to outsource the manufacture to the third parties to reduce the cost. The rapid iteration and updating of IC technologies have led many key designers to outsource the manufacturing process to third parties. To this day, many design firms have cancelled manufacturing plants driven by lots of motivations including reducing cost, focusing

This work is supported in part by the XXX.

Jianfeng Wang, Zhonghao Chen, Jiahao Zhang, Enze Ye, Ziheng Zheng, Huazhong Yang, Yongpan Liu and Xueqing Li are with BNRist/ICFC, The Department of Electronic Engineering, Tsinghua University, Beijing 100084, China (e-mail: wangjf18@mails.tsinghua.edu.cn, chenzh19@mails.tsinghua.edu.cn, Jiahao-z19@mails.tsinghua.edu.cn, zhengzh18@mails.tsinghua.edu.cn, yez19@mails.tsinghua.edu.cn, yanghz@tsinghua.edu.cn, ypliu@tsinghua.edu.cn, xueqingli@tsinghua.edu.cn)

Yixin Xu, Tongguang Yu and Vijaykrishnan Narayanan are with the Department of Computer Science and Engineering, Penn State University, University Park, PA 16802, USA (e-mail: ybx5145@psu.edu, tvy5113@psu.edu, vxn9@psu.edu)

Sumitha George is with the College of Engineering, North Dakota State University, Fargo, North Dakota state, USA (e-mail: sumitha.george@ndsu.edu)

Digital Object Identifier XXXX/XXXX.XXXX

on core design and capabilities and enlarging the worldwide

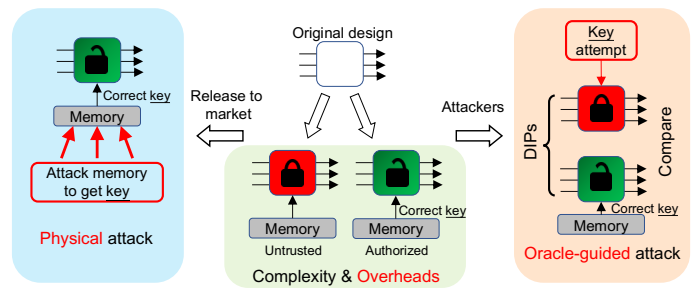

Fig. 1. Overview of the challenges of key in logic locking: physically attack the key from memory; convenient key traversal attempts to establish the oracle-based attack; the trade-off between the key obfuscation effect and overheads.

market [1].

Outsourced manufacturing can lead to situations where IC supply chain can be long and worldwide-distributed, in which third parties may be untrusted, resulting in problems of hardware security inevitably. The global trends in IC supply chain, including IC design, synthesis, verification, fabrication, testing, packing and system integration, make it possible for the third parties to operate abnormal process. IP (intellectual property) piracy is a critical threat since the attackers may reverse engineer the chip to obtain the design details, and the counterfeiting can lead to the profits which are not belongs to the IP designers [2]. To mitigate the threat, several design-for-trust (DfTr) methods have been proposed. Mainstream countermeasures include IC metering [3][4], watermarking [5][6], split manufacturing [7], IC camouflaging [8] and logic locking [9]–[12].

Among all the countermeasures, logic locking is a wild-accepted gate-level technique by inserting additional logic gates controlled by the key-inputs (key) which is assumed to be stored into a non-volatile memory after fabrication by the IP vendor [C6], and then the outputs are exactly correct only when the key is correct. While logic locking has a greater protection range in the manufacturing flow, it suffers from oracle-based attacks [13][14] (especially SAT attack, one type of algorithm attack to decipher the key [13]). SAT attack uses distinguished input patterns (DIPs) to examine the possible key by comparing the computing results of two circuits: an oracle which already stores the correct key and a chip to be attacked. Since the two chips have the same DIP input, whenever the key entered by the locked chip to be attacked is wrong, the computing result will not match that of the oracle, so the key must be wrong and deleted. The attacker continuously tries different key on the chip to be attacked, so as to quickly delete all the wrong key and finally obtain the correct key.

As shown in Fig. 1, the security and obfuscation effect of logic locking totally rely on the key (i.e., the location of key-gates and the key length), however, current attack and defense works treat the key attempting as a very simple matter. The key is considered to be stored in memory, and the attacker's attempt can be thought of as arbitrary modification of the key through the interface or scan access between the memory and the computing circuit. However, at least there are three fundamental flaws in this key-related assumption. Firstly, it is a simplistic assumption that the key can be easily modified for attack attempts, but in fact the scan chain can be easily blocked by a secure scan interface at low cost to prevent attempts at the key at will [15]. The attacker can attempt a key, but it may take more time to obtain a desired key and the attack process will be prolonged. Secondly, the correct key is stored in the memory on the oracle chip, and the security of the storage unit was questioned [16][17]. Once a hacker has physically cracked the right key, logical locking, no matter how effective, will fail [18]. Thirdly, the key length is a critical variable for security. Existing protection measures usually add redundant protection circuits to reduce the pruning speed of hackers in algorithmic attacks, resulting in the low error rate and more overheads. The increase of key length (often proportional to overheads) should be a means to increase obfuscation effect, but letting the error rate go down to avoid pruning is a step back [19].

This arbitrary modification of the key also weakens the integrity of the key. Since the difficulty of overall cracking is significantly greater than the difficulty of cracking part by part after decomposition, the hacker can decompose the key corresponding to the circuit module, so as to carry out the traversal. Current major logic locking is mainly focused on locking a specific circuit module instead of the whole chip. Therefore, hackers can remove each module from different IP vendors to attack separately. Prior works (i.e. sequential de-obfuscation) also discuss the situation of closing the scan access so as to combine the key bits as a whole [20], but they are not as effective as SAT attacks with respect to time cost. Therefore, both SAT attacks and sequential de-obfuscation algorithms are not applicable to unlock today's VLSI circuits, if we find an approach to locking the chip as a whole instead of locking each module offered by different IP vendors respectively.

Module-level sequential obfuscation makes key acquisition and traversal difficult, but state complexity becomes a new problem [21]. This work utilizes the complex state space of a CPU core, in which the internal nodes' possible value combinations (invisible to hackers due to secure scan interface) form a complex state transition graph (STG) and the states are transferred by an input instruction sequence (IIS). In the huge state space, very few states are specifically chosen to be the correct states, which represent the key. The exponential mapping between the IIS and the key provides a huge exploration space, and the reverse solving of the IIS from key are extremely difficult. Note that this work only reuses the original large FSM in a CPU core to generate the key, this method has a large state number and does not need to store the key explicitly in memory, which mitigates the challenge mentioned before.

To further enhance the defensive effect, recently emerging Embedded Programable Logic (EPL) method is considered to be a good way to hide the sensitive parts of the circuit designs [22]. By adding the reconfigurable devices/blocks/modules into the original design, the circuit behavior of the same layout can be various controlled by different key. Therefore, the attacker cannot obtain the design from the layout by reverse engineering. The only pity is that the EPL, as well as the logic locking and IC camouflaging, has the overheads that cannot be ignored, and the emerging gate camouflaging using threshold voltage defined (TVD) is proposed to mitigate the issue [23]. TVD hides the information of different logic functions in the same topology by utilizing the programmable threshold voltage ($V_T$). TVD can be implemented by various of devices, and this work adopts ferroelectric FET (FeFET) mentioned in [23] since it is compatible with CMOS logic technology.

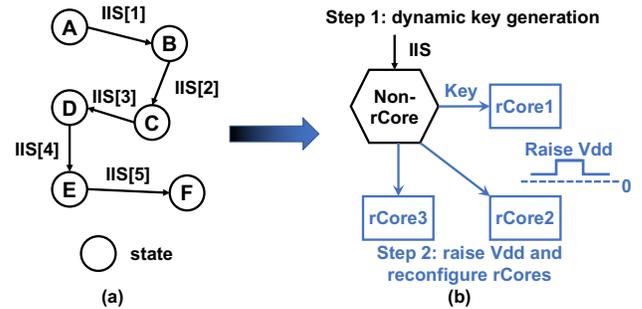

**Fig. 2.** Illustration of using FSM to generate the state as key: (a) State transition diagram of 5 IIS; (b) Two steps to configure rCores using the key generated by a non-rCore.

Inspired by the above thoughts, this work proposes a new logic locking method, ALL-MASK, to achieve a more comprehensive hardware protection effect at the lower cost of area, power consumption and delay. Shown in Fig. 2, this work relies on the existing enormous FSM to protect the original design, in which the customized IIS are used to set the FSM to the correct state. The target state is reused as key to configure the reconfigurable cores (rCores) to the target functions. The huge attack space of the STG and the TVD topology provide a thorough protection from the global architecture design to the local layouts and gates. In the reconfiguration process, it is also time-consuming process to raise the *Vdd* for writing FeFETs, which also improves the security of cracking

In Fig. 3, we demonstrate the secure method of locking and unlocking the chip. Fig. 3(a) shows that our encryption is an on-chip level encryption that does not require additional configuration ports and key storage units. From an outside perspective, the user can only attempt by simply running the IIS and raising the power voltage *Vdd* for a period of time. There is no extra port or additional memory to attempt or traverse the key. Fig. 3(b) shows the detailed steps. The original chip design is a multicore architecture, and we lock (n-1) cores and choose one core to generate key. We reuse the signals in that core, without storing the key and adding additional configuration ports. When the correct key bits are generated, all of the logic locks they controls can be unlocked by raising the voltage for a period of time. If the key bits do not unlock all logical locks, the unlock still fails. It is worth noting that the number of logic gates inside cores is much larger than the number of encrypted logic gates, so extra costs are very low.

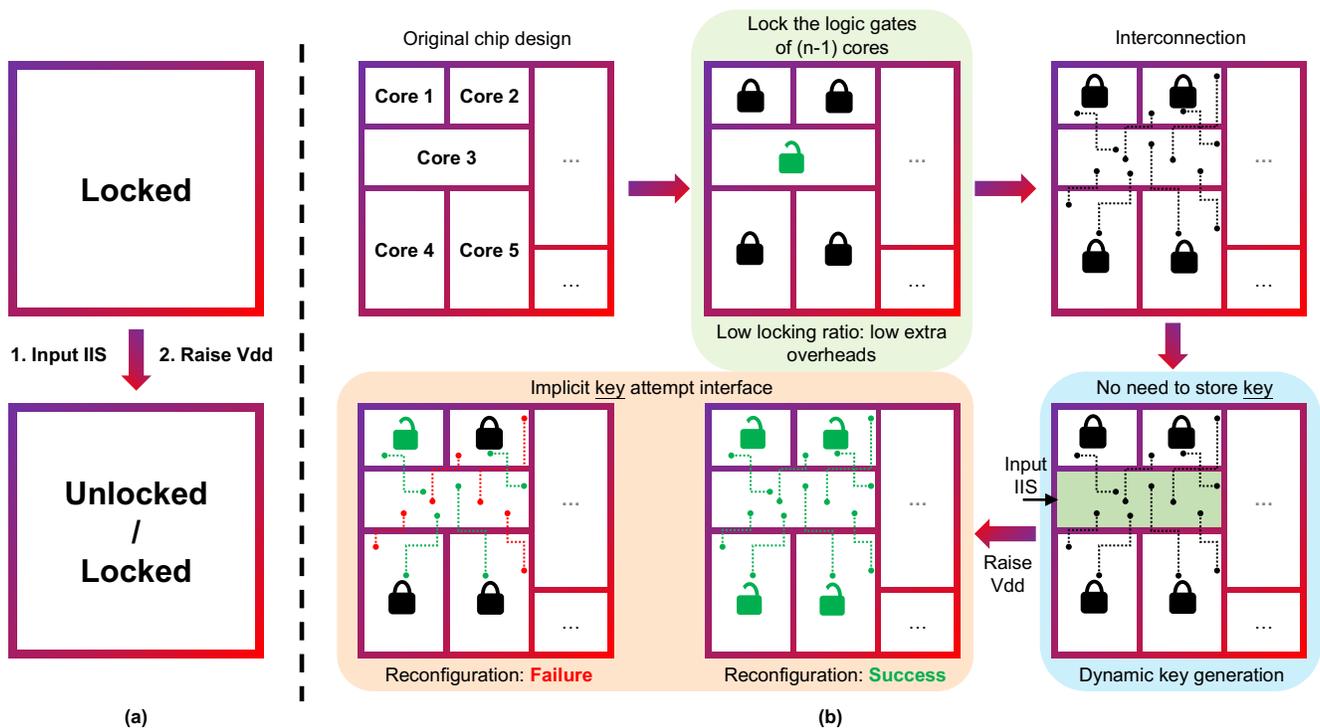

**Fig. 3.** Proposed ALL-MASK steps to lock and unlock a multi-core chip. (a) Black box perspective. (b) Detailed steps of locking the cores, connecting the key bits and locked gates, generating key dynamically and unlocking the cores.

This work makes the following contributions:

1) *ALL-MASK architecture.* This work proposes a hardware secure method design for a multicore processor and lock the processor as a whole instead of several separated modules. The method reuses the original complex FSM of a non-rCore and utilizes the low overhead gate-level camouflage method TVD to achieve a comprehensive protection from the global design to the local layout. By appropriately inserting FeFETs in non-critical paths, our design will not have a negative impact on clock frequency. In this work, one core is used to encrypt the other cores, and the configuration is carried out inside the chip, eliminating the explicit configuration information flow, which is a covert encryption configuration means.

2) *Novel dynamic key operating strategy.* Firstly, this work considers, evaluates and predicts the difficulty of obtaining key during state transition through the scan-blocked key attempting. Because this work takes advantage of the feature that all reconfigurable gates must be configured at the same time, all key must meet the correct conditions at the same time to be successfully configured. Secondly, non-rCore possesses a huge key-searching space searching by inputting IIS makes the attack extremely hard does not require extra memory to store the key. No extra input and output ports makes the method imperceptibly and light-weighted.

3) *FeFET-based reconfigurable gate (rGate).* This work proposes a method of constructing reconfigurable logic gates (rGates) using TVD devices, and recently emerged non-volatile devices (e.g., RRAM, FeFET) are good alternatives. To the best of our knowledge, we build the FeFET-based reconfigurable ALL-MASK architecture using IIS to generate key, and propose a FeFET-based gate-level obfuscation method. This FeFET-based logic locking method is designed especially for the gate-level logic obfuscation for its convenience, low cost and good logic scalability. Since FeFET writes process are typically in the nanosecond to microsecond range, this time cost is a major deterrent to cracking. Considering the astronomical number of transistors in today's VLSI circuits, the overhead (delay, area, power consumption) of introducing FeFETs is negligible.

In the rest of the paper, section II introduces the prior work. Section III introduces the FeFET-based ALL-MASK method. Section IV presents the analysis and evaluation. Section V concludes the work.

## II. BACKGROUND AND MOTIVATION

### A. Prior Work and Our Insights

Existing hardware protection methods can be broadly divided into two main categories: obfuscation-based protection and authentication-based protection [24].

*Obfuscation-based protection*: obfuscation-based measures relies on the specific encryption techniques to hide the core design on a certain level, and current mainstream methods include IC metering [3][4] (FSM/HDL level), logic locking [9]–[12] (gate level) and IC camouflaging [8] (layout level). IC metering provides a way for the chip designers to obtain post fabrication control by adding an FSM with massive incorrect state to hide the correct state. Emerging FSM reverse engineering provides a way to reconstruct the high level design [25], and we notice that FSM obfuscation method is limited to the degree and number of the logic and state, respectively [21], and the limited state number makes traversal and inference easier for the attacker.

To further enhance the security and mitigate the potential threat brought by software and program language, hardware obfuscation method emerged to hide the critical design with peripheral structure, which usually does not change the original function. Logic locking and IC camouflaging are typical techniques of hardware obfuscation. Logic locking inserts additional logic into the target logic module, and the function depends on primary inputs and key inputs. Key inputs are fetched from the off-chip memory and sent to the circuit module through input ports. Any untrusted party without the correct key cannot obtain the correct function [9]. IC camouflaging is a layout-level technique and use camouflaging cells to synthesize the circuit. The camouflaging cells look alike but have different functions, thus the original design cannot be determined from reverse engineering [8].

While the above two methods provide good hardware protection and many related techniques are proposed [8][22][23], they still encounter three possible problems: oracle-guided attack (mainly SAT attack [13][28]), physical attack [16][17][18] and large overheads [8][29]. Oracle-based attack is an outstanding threat to logic locking and other obfuscation methods. Assuming the hacker has an oracle, which is a chip that has already store the correct key into the memory, he can input the same primary inputs and different key inputs into both the locked chip and the oracle. By comparing the outputs, the wrong key can be eliminated when the two outputs are different. This pruning process can be finished in a short period of time and thus cracking the locked IC. SAT is the most widely used oracle-based attack to guess the correct key by taking advantage of Boolean satisfiability-based key-pruning algorithm to find distinguishing input patterns (DIPs) to prune the wrong key in a short time [13], which is mainly designed to attack combinational logic circuits [20]. Most SAT attacks implicitly assume that the internal nodes of chips can be monitored and modified via the scan chain [30], which enables hacker to decipher the key easier. However, since the user-mode access to the scan chain can be blocked by a secure scan interface at low cost [15], SAT attack will defunct under this circumstance [17][27]. There do exists other oracle-based attacks without access to scan chain [31], which are also called sequential de-obfuscation algorithms, but they are not as effective as SAT attacks with respect to time cost [32]. Most of the proposed defense methods are against oracle-based attack, such as SFLL [9], Anti-SAT [10], SARLock [11] and TTLock [12]. However, these works might be more vulnerable to other attacks (e.g., removal attack [33]) or reduce obfuscation effect (e.g., Anti-SAT will decrease the error rate [19]). Worse still, many novel oracle-based attacks have been proposed to crack these methods. Worse still, many novel attacks have been proposed to crack these SAT-resilient methods, such as Double DIP [34], AppSAT [14], bypass attack [35] and removal attack [36].

In addition to oracle-based attack, physical attack is another emerging topic. Major anti-SAT methods store the key into the non-volatile memory unit, yet the security of the memory is remained unsure. Traditional logic locking assumes a tamper- and read-proof secure memory to store the key and ignore the risk of key extraction when the user transfers the key between the memory and the locked circuit [16]. Likewise, major SAT-resilient methods store the key into the non-volatile memory unit. Nevertheless, the security of the memory is remained unsure and the process of key transfer is susceptible to malicious attacks. The key can be physically revealed through invasive attacks, and [18] summarizes the previous work of the potential physical attacks to obtain the key. [16] demonstrated an experiment on FPGA that successfully extracts the secret key during the process of key transfer using optical probing. [37] exploited power side-channel analysis to launch SAT attack. Besides, beyond the fact that key may be gained from the memory unit by some means without implementing pruning algorithms, this directly access and modification also makes it easy for hackers to modify and replace the key during the hacking attempt. Current SAT attack are based on the key-pruning process, and various key combination attempts are usually considered to be easy and available.

Logic locking and IC camouflaging also face the tradeoff between the key length overheads, summarized in [8], [29], respectively. The key length is usually related to the key inputs, which is proportional to the extra latency, power consumption and area caused by the peripheral protection circuit. Reducing the key length to reduce the overheads is not a good choice since it mitigates the protection effect against the algorithm attack. We note that these large peripheral structures brought by large key bits not only generate high overheads, but also are a prime target for reverse engineers (e.g., removal attacks [9]). To address the issue, we attempt to construct a reconfigurable logic architecture ALL-MASK (introduced in section III) using embedded programmable logic (EPL) and low-overheads non-volatile (NV) device (introduced in section II.B), targeting to achieve a lightweight protection method.

*Authentic-based protection*: authentic-based methods embedded the digital watermarks into the circuit design to prove the identity of the designer or manufacturing party. The watermark cannot be removed artificially by some technical means, and thus can be used as an IP proof [5]. Despite providing a protection on IP, authentic-based methods is limited due to the passiveness while facing the attacks, and the stolen/fake IPs can still function well [24].

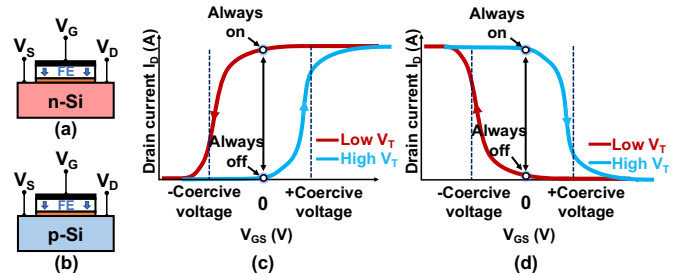

**Fig. 4.** FeFET adopted in this work. (a) N-type FeFET device structure. (b) P-type FeFET device structure. (c) N-type FeFET I-V curve. (d) P-type FeFET I-V curve.

## B. Non-volatile device and FeFET

Emerging non-volatile (NV) devices have shown excellent develop prospects in computing [38], memory [39] and energy harvesting applications [40]. NV devices exhibit different electrical characteristics by switching internal state, which will not loss when the power is cut off. The reconfigurability of internal state provides a new camouflage way to program after the manufacturing process, which effectively prevent the reverse engineering (RE). Threshold voltage defined (TVD) logic topology is a fundamental idea that utilizes the difference electrical behavior of different threshold voltage ($V_T$) to hide the original design [41][42].

FeFET is an NV device that embeds a ferroelectric (Fe) capacitor into the transistor stack (Fig. 4(a-b)). With the use of HfO2-based Zr-doped material, FeFET enables a scalable and CMOS compatible ability to keep the pace with leading-edge logic technologies [43][44]. FeFET is a TVD device [23] and has a reconfigurable $V_T$. This work adopts two types of FeFET: n-type and p-type, which are evaluated in [45], and in this work the both types are considered to store a single-bit binary state: always-on and always-off. The two states can switch by applying $V_{GS}$ beyond the coercive voltage, and the I-V curves are shown in Fig. 4(c-d).

FeFET has a write endurance about $10^5$-$10^{10}$, depending on the device optimization and configuration process [46]. This limited endurance also provides a guarantee to restrict the hacking and attempting times.

## III. METHODOLOGY OF ALL-MASK

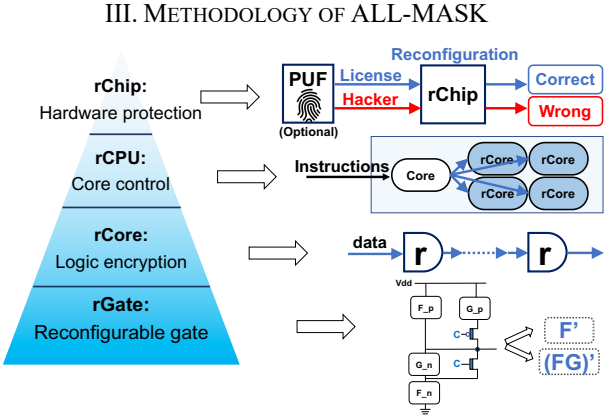

**Fig. 5.** ALL-MASK design levels: rGate, rCore, rCPU and rChip.

This section introduces the ALL-MASK and presents a way to secure a chip. The proposed bottom-to-top approach, as shown in Fig. 5, combining from the gate level to the application level. ALL-MASK deploys reconfigurable gates (rGates) to implement the programmable logic. The rGates are used in a CPU core to construct a reconfigurable core (rCore), and a number of rCores are controlled by a normal CPU core, which constructs a multi-core reconfigurable CPU (rCPU). Furthermore, with the rCPU as a core design, extra authentication units such as PUF can be added to construct a reconfigurable chip(rChip) to get extra protection.

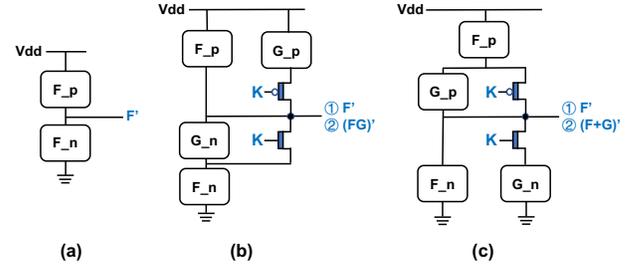

**Fig. 6.** Basic circuitry to construct the reconfigurable micro-architecture. (a) Original CMOS logic gate. (b) Proposed type-1 rGate. (c) Proposed type-2 rGate.

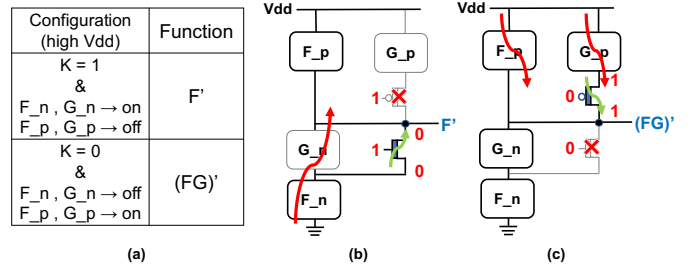

**Fig. 7.** Method of configuring an rGate. (a) Reconfigure table. (b) Reconfigure to F'. (c) Reconfigure to (FG)'.

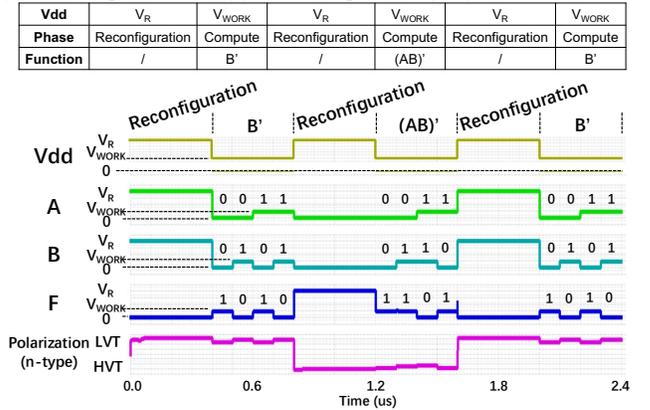

**Fig. 8.** Waveform of the rGate function simulation.

### A. Reconfigurable Non-volatile Gate

Gate-level rGate is considered to be the fundamental unit of implementing ALL-MASK. This work proposes two types of rGate and method to construct them. As shown in Fig. 6(a), the original CMOS logic gate functions is F'. The type-1 rGate and type-2 rGate utilize an additional logic $G$ to obfuscate the original logic. As shown in Fig. 6(b-c), type-1 and type-2 rGates function as {F'/(FG)'} and {F'/(F+G)'}, respectively, and the reverse engineering cannot reveal the correct logic due to the FeFET TVD characteristic. In the following, we take type-1 as an example to introduce the two modes of the circuit behavior: *reconfiguring mode* and *computing mode*.

*Reconfiguring mode*: reconfiguring mode is designed to change the functionality of the rGate. In this mode, $Vdd$ is set to $V_R$, a value that beyond the coercive voltage to provide the enough $V_{GS}$ for changing the inner state (polarization state/$V_T$) of the FeFET. Fig. 7(a) explains the configure process. In Fig. 7(b), if the $F\_n$ and $G\_n$ are both on and the $F\_p$ and $G\_p$ are both off, for n-type FeFET, the source and drain are grounded; for p-type FeFET, the source is floating and drain is grounded. If the voltage of $K$ is $V_R$, according to Fig. 2, the n-type and p-type FeFET are reconfigured to low-$V_T$ (LVT) state, which represents "always on" and "always off" behavior, respectively. Thus, the rGate functions as $F'$, since the $G\_n$ is short-circuited and the branch where the $G\_p$ resides is open-circuited.

In Fig. 7(c), if the $F\_n$ and $G\_n$ are both off and the $F\_p$ and $G\_p$ are both on, for n-type FeFET, the source is floating and the drain voltage is $V_R$; for p-type FeFET, the source and drain voltage are both $V_R$. If the voltage of $K$ is $0$, according to Fig. 2, the n-type and p-type FeFET are reconfigured to high-$V_T$ (HVT) state, which represents "always off" and "always on" behavior, respectively. Thus, the rGate functions as $(FG)'$, since the $G\_n$ is not short-circuited and the branch where the $G\_p$ resides is not open-circuited.

*Computing mode*: computing mode is designed for normal computing process with a stable function for each rGates. In this mode, $Vdd$ is kept stable at $V_{WORK}$, which is lower than the coercive voltage and promises the circuit can work as a normal CMOS circuit. The polarization states of FeFETs will not change and are consistent with the state configured by reconfiguring mode.

The proposed novel rGate structure has significant advantages in hardware obfuscation. Unlike other FeFET-based reconfigurable gates which focus on multi-logical functionality [23], this design can easily modify the original CMOS logic gate into a new rGate since the $F$ and $G$ can be random combination logic. The replacement principle from CMOS gate to rGate is introduced in section III.B. An example of a type-1 rGate $\{B'/(AB)'\}$ is simulated and the waveform is shown in Fig. 8. The reconfiguring mode changes the polarization state of FeFETs, and the working mode achieves two different functions: $B'$ and $(AB)'$.

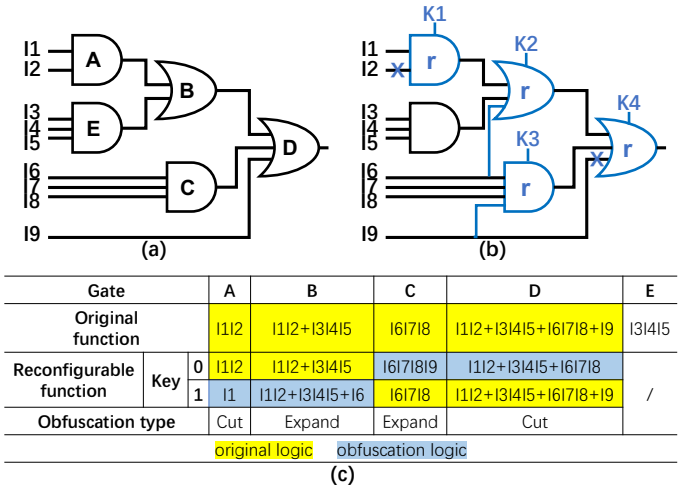

**Fig. 9.** Proposed four types of gate-level replacement to construct an rCore. (a) Original CMOS logic module. (b) Logic module with rGates. (c) Replacement type table.

### B. Reconfigurable Non-volatile Core

ALL-MASK provides four types of gate-level obfuscation to replace the original logic gates into rGates, constructing a reconfigurable Core (rCore). Fig. 9 depicts the replacement methodology. Fig. 9(a) shows an original CMOS logic module, and Fig. 9(b-c) shows the corresponding replacement policy.

Type-A: type-A applies to "AND gate" and the obfuscation logic is "cut" down from the original logic. As shown in gate A, the original logic is $I1I2$, equivalent to "$FG$" in two logics of type-1 rGate, and the other logic can be selected as "$F$", which is I1 in this case.

Type-B: type-B applies to "OR gate" and the obfuscation logic is "expanded" from the original logic. As shown in gate B, the original logic is $I1I2+I3I4I5$, equivalent to "$F$" in two logics of type-2 rGate, and the other logic can be selected as "$F+G$", which is $I1I2+I3I4I5+I6$ in this case.

Type-C: type-C applies to "AND gate" and the obfuscation logic is "expanded" from the original logic. As shown in gate C, the original logic is $I6I7I8$, equivalent to "$F$" in two logics of type-1 rGate, and the other logic can be selected as "$FG$", which is $I6I7I8I9$ in this case.

Type-D: type-D applies to "OR gate" and the obfuscation logic is "cut" from the original logic. As shown in gate D, the original logic is $I1I2+I3I4I5+I6I7I8+I9$, equivalent to "$F+G$" in two logics of type-2 rGate, and the other logic can be selected as "$F$", which is $I1I2+I3I4I5+I6I7I8$ in this case.

The logic obfuscation can be used in the selected data path and control path as long as the logic gates take part in logic computing. The output is then affected by the rGates and the key are critical to ensure a fully correct function. The obfuscation effect and the hamming distance (HD) between the correct outputs and obfuscated outputs are evaluated in section IV.C, and the latency, power consumption and area overheads are evaluated in section IV.D. It should be noticed that this rGate-replacement methodology is not limited to FeFET-based rGate, any other devices that has the similar characteristic can be applied to this method to enlarge the design space.

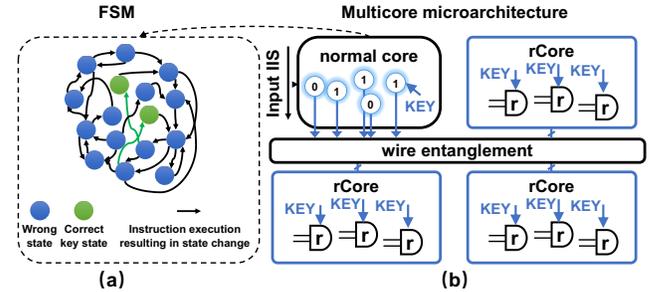

**Fig. 10.** (a) Normal core FSM. (b) Structural obfuscation scheme by using one normal core (non-rCore) to set other rCores states.

### C. Microarchitecture Co-design

The microarchitecture co-design is mainly focused on the software-hardware coordinating protection. Section III.A and section III.B are mainly targeting at hardware level to utilize the structural characteristic to protect the original design. Here, ALL-MASK enhance the protection at software level by leveraging the mismatch between the circuit nodes and input instruction sequence (IIS) in FSM obfuscation. HARPOON [24] proposed this type of design methodology with small-FSM attempt, and we intend to enlarge the FSM to a new level by reusing the FSM of a CPU core. The giant number of states are constructed by the sequence of numerous of node digital values, and the selected nodes can be used as key to unlock the rGates.

The basic strategy is shown in Fig. 10. ALL-MASK uses the FSM of a core which do not contains any rGates to generate key. This FSM is certain since there are no rGates to provide obfuscation in the normal core. As shown in Fig. 10(a), the normal core behavior is described by an FSM. The states are the combination sequence of the nodes value, and the IIS is the direct interface for state transition. Among the massive states in the FSM, only an extremely small part can guarantee that all the selected nodes value are correct simultaneously. The IIS shifts the state of the normal core to the correct key state, making the key correct at the same time. The IIS is determined by the hardware structure and the ISA of the normal core, which is designed by the designer in advance. Fig. 10(b) describes the control relationship between the key and the rCores. The key generated with the state are connected to control the rGates in rCores to set the functionality.

TABLE I
QUALITATIVE ANALYSIS OF ALL-MASK COMPARED WITH PRIOR WORKS.

| Major challenges | Causes | Our insights | ALL-MASK advantages |
|---|---|---|---|
| Physical attack | Memory stores key | Cancel the storage part | Key inputs are generated by a core with IIS without specific memory to store key |
| Easy key attempts and attacks | Explicit key interface | Embedded Programable Logic (reconfigurable circuit) | Implicit reconfiguration between the cores inside the chip |
| | Separated circuit modules | | Configure the modules at the same time |
| Key-related overheads | Large peripheral circuit related to the key length | FeFET characteristics | FeFET-based rGates with extra-low overheads |

FSM is the behavior description of the normal core, and it converts the IIS to the key. Fig. 11 provides an example to explain the process about an 8-bit key generating. IIS is designed ahead and input to the normal core orderly. The FSM, which transits the states according to the instruction in IIS, is extremely big due to the giant number of nodes. The pre-selected nodes form a core part of the FSM, since only changes in this part involve transition to correct key. Starting from the "reset" state, the selected sequence of node values begins state transition through IIS input, and finally enters the target state (green circle), where all of the key bits are correct simultaneously. Thus, a target 8-bit key {00001001} is generated through the specific IIS.

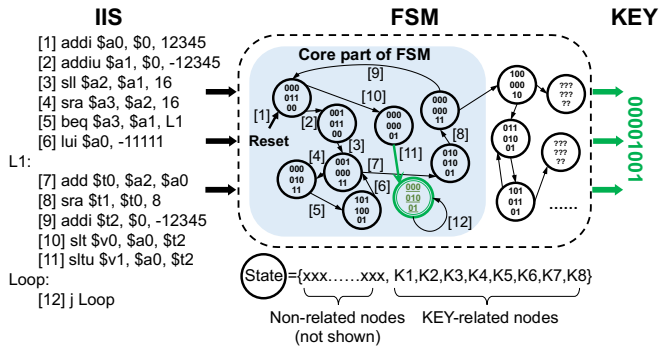

**Fig. 11.** An example on how to generate key through input IIS.

The utilizing of the normal core to generate key provides two advantages to enhance the security. On the one hand, as shown in Fig. 11, the relationship between the IIS and the key are complex due to the huge FSM. Even if the structure of the normal core is reverse engineered by the hacker, which means the FSM is possibly leaking to hackers, the reverse solving from the key to the IIS is still a very difficult challenge. On the other hand, the key bits are combined as an entirety and the changing process is an overall procedure. During the transition of the FSM, the attempt to change a small part of the key bits is likely to result in replacement of the rest of the key bits. The bit-by-bit modification is hard to be implemented in the FSM, and such design plan provides much higher resistance to the algorithm attack which focus on key pruning.

The location choice of the nodes is a topic in real-time systems worthy of further discussion in the future. Temporal systems involve very complex state transitions and cumulative effects. The result of each calculation is not only related to the input at the current moment, but also likely related to the previous calculation result or process. Location selection has great design freedom and space, and the basic principle is to make the bits as relevant and dependent as possible. The dependency prevents the bit-by-bit attacking attempt and it is difficult to construct a certain possible combination sequence of key. [47] discusses the formal verification of the real-time systems and [24] proposes a suitability metric for the node chosen and evaluation standard.

The prior works have presented several strategies and methods for location selection which can be used in ALL-MASK, and we also focus on the number of bits of the key since the bitlength is a critical variable in hacking and obfuscation. Although a certain number of states can lead a particular sequence of key to the correct state, increasing the key length greatly reduces the number of correct states. In other words, letting more nodes to reach the correct preset voltage at the same time is a harder constraint, and the hackers need to satisfy a much harder condition with a difficult IIS design process.

With the choice of IIS and nodes, the key sequence can be seen as the configuration bits and the rGates take the key to reconfigure the functionality. As depicted before, the $V_{dd}$ needs to be raised to $V_R$ and lasted for a period of time for the FeFETs to change the polarization state, which is a stealthy and low-overheads process. Additionally, take rCPU as the core IP, the designers can add a variety of peripheral modules for more comprehensive identity authentication to build a reconfigurable Chip (rChip, Fig. 5). For example, a physically unclonable function (PUF) can provides verification of $V_{dd}$ controlling to prevent the $V_{dd}$ being uncontrollable. With a stable $V_{dd}$, the FeFETs states cannot be changed easily, which protects the system from being maliciously implemented by Trojans.

IV. ANALYSIS AND EVALUATION

*A. Attack Model*

As aforementioned, we propose a FeFET-based multicore architecture to ensure security. Here is the attack model.
Threats:
1. Hackers can have access to the whole netlist and layout of

the chip through the global IC design flow.

2. Hackers can buy an unlocked chip from the market to launch oracle-based attacks.

3. The memory used for key storage is not tamper- and read-proof and the key may be extracted by exploiting physical attacks.

Constraints:

1. Hacker cannot have access to the secure scan interface to check or modify the internal nodes during runtime.

2. Hacker cannot get the IIS for correct configuration offered by the IP vendor and have to decipher the key via algorithm attacks or physical attacks.

3. Hacker has to configure the whole FeFETs to the right polarization states and cannot decipher each module from different IP vendors respectively due to our design.

The attack model in this work assumes that the hacker has an unlocked (activated) chip, which has already generated the key by input the correct IIS. With the correct key link to the rGates simultaneously, the configuration process reconfigures the internal states of the FeFETs and do not require the specific memory to store the key. The unlocked chip is considered as an oracle, and the correct output can be obtained through continuous questioning (by input arbitrary primary inputs and key inputs). With the oracle, the attacker has the locked chip and the relevant reverse-engineered locked netlist. In this work, the inner states cannot be accessed via scan chains, so that the questioning is only rely on IIS and output computing results. The attacker has the access of input and is able to use IIS to set the inner state. After the configuration, the attacker examines the rCores by performing the logical calculations. The SAT-based algorithm attack rules out the incorrect IIS and inner states by comparing the computing results of the rCores between the oracle and the locked chip. Once the two results of the two chips are different, the current IIS and inner states of the locked chip cannot be correct. The hacker has to change the IIS to update the inner states of the normal core and reconfigure the rCores, and the pruning process is the same.

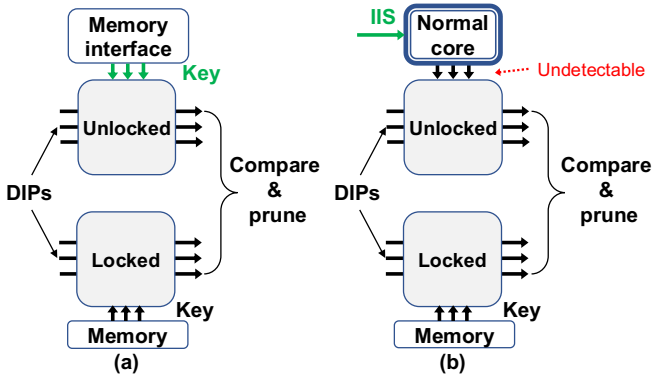

**Fig. 12.** Attack model. (a) Common SAT-based attack model. (b) SAT-based attack model for ALL-MASK.

Fig. 12(a) shows the common attack model of SAT. By repeatedly comparing the results while traversing the key inputs, the wrong key can be quickly eliminated in a short period of time. ALL-MASK modifies the key-generated process by using a normal core, and the SAT-launching process needs the key inputs to generate by IIS. As shown in Fig. 12(b), the locked chip using ALL-MASK method requires key inputs generated by IIS instead of input directly from the memory-compute interface.

## B. Qualitative Analysis and Comparison

This section qualitatively analyzes the advantages of ALL-MASK compared to the previous obfuscation-based hardware protection methods. As shown in Table I, the major challenges brought by the corresponding countermeasure is introduced briefly in section I and II. This work combines the advantages of prior works, and innovatively puts forward dynamic key generation and FeFET-based hardware protection methods to mitigate the challenges.

Algorithm attack (e.g., SAT attack) is the critical issue in the logic obfuscation. Fig. 13 (a-b) explains the two enhancements of ALL-MASK. Firstly, traditional defense methods focus on hardware modifications bas ed on the mathematical relation of key inputs and DIPs. We notice that the default key selection and traversal is an ideal approach, which provides traversal samples for the attacker to test and compare results. Instead of interacting key directly from the memory, ALL-MASK uses a normal CPU core (without rGates) to generate a set of key. As introduced in section III.C, key generation is not simple and requires state transition in FSM. In other words, it is difficult to obtain a specific key sequence since FSM restrictions prevent such arbitrariness. Secondly, the prior works often focus on the module level protection, which allow the hackers to crack each protection peripheral module separately. ALL-MASK leverages FSM to combine the key as a whole, and the rGates are configured at the same time. The simultaneously key-changing and the gate-reconfiguration force the idea of reducing complexity by separate hacking invalid. The theoretical analysis and experimental results are shown in section IV.C to support the analysis.

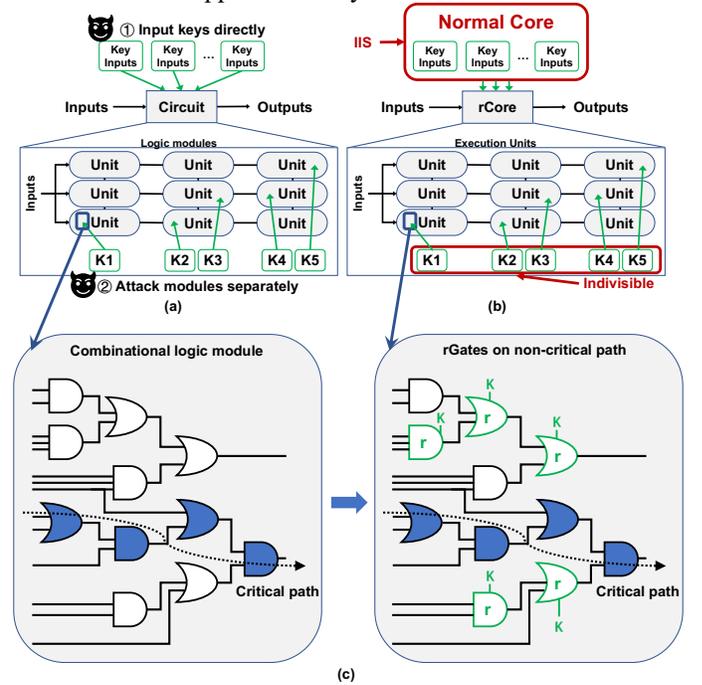

**Fig. 13.** (a) Two important conditions for separate key. (b) ALL-MASK structure to enhance the algorithm attack resiliency. (c) Replacement policy.

Prior works based on logic locking or IC camouflaging are also prone to suffer from physical attack, since the traditional logic-obfuscation methods require a specific memory unit to store the key, and the memory unit becomes a potential attack site. ALL-MASK uses a normal core to generate the key without extra memory unit, and the control of key requires a real-time interaction through IIS. This key-from-compute process provides a lightweight key protection approach, and the key length can be very long to enhance the security without the concern of storage.

Non-negligible overheads (latency, power consumption and area) are challenges to the existing protection methods caused by the significant hardware overheads, since they expected to resilient various potential attacks with fixed hardware peripheral circuits. The main bottleneck of the existing peripheral circuits is the tradeoff of key length and protection effect. Peripheral circuits scales usually control outputs with primary inputs and key inputs, resulting in a bottleneck of the key length. In this work, ALL-MASK makes very minor changes to the logic gates by constructing rGates. The modifications can obfuscate the functionalities while keeping them lightweight at low overheads, which significantly boosts the upper limit of the key length. The simulation results and analysis are shown in section IV.E. Apart from the gate-level obfuscation, the TVD characteristic of the FeFET effectively defends reverse engineering, hiding the critical information of the original design.

To further improve the security, the design idea of hardware and software co-design is adopted in the prior related works. FSM-based obfuscation has been a hot topic and there's a lot of space for improvement in the number and complexity of the states. ALL-MASK takes advantage of the characteristics of the normal core as a complex FSM, multiplexing node voltage as configuration stream (key inputs) on the basis of the FSM. With the large original FSM as the foundation of obfuscation, the states corresponded to the selected key occupy a very small proportion in the whole map. The transition controlled by the IIS further reduces the probability of reaching these key states. In addition, reusing the state machine not only increases the state space, but also makes the state-adding process no longer explicit, which improves the ability of code and obfuscation concealment.

*C. FSM and Key-Traversal Analysis*

As mentioned above, the states of rGates (i.e., polarization states of FeFETs) in rCores consist of each bit of the key. To enhance efficiency and reduce overheads, as shown in Fig.10 and Fig.11, we utilize a normal core(non-rCore) to generate the correct key indirectly by using a specific IIS. Assume that there exists S nodes in the non-rCore, among which we elaborately select K nodes as key. These K nodes are connected to rGates in rCores. Only when the K-bit sequence accords with the originally set key, can we obtain the correct functionality of the designed circuit. The huge space of the normal core's FSM effectively ensure security. Obviously, there are $2^S$ states and only $2^{S-K}$ of which are correct ones. It is noteworthy to note that the proportion of the correct states is only $2^{-K}$, which decreases exponentially which the linear growth of key-length. Due to the FSM obfuscation design of the non-rCore as well as the inner configuration mechanism of rGates in rCores, the hacker cannot prune the key and has to decipher the correct key by running IIS instead of modifying the key inputs directly.

Each time the hacker uses an instruction to reach a possible state of the key, reconfigures the rGates in rCores and then tests functionality of rCores, which increase the time cost of each key attempt. Therefore, it is much more difficult for the hackers to get the correct key. We will demonstrate the theoretical analysis inspired by fetching-ball model in Classic Probability Theory and several simulation results to support our insights.

We introduce a probability model to further discuss traverse complexity of deciphering the key. Since the key are composed of several internal nodes of the non-rCore, the hacker has to obtain the correct key by running IIS instead of modifying the key directly. For each time the hacker chooses a possible state of the key and then reconfigures the rGates in rCores, he can decipher the key only by reaching a correct state. all model in classic probability to describe the problem.

This work uses mean value to describe the traverse complexity since the ultra-low probability of getting the correct key within a short period of time is negligible (which is shown in Fig.14). Some notations are as follow.

·$K_1, K_2, ... K_m$: Elaborately chosen $m$ bits in the non-rCore. Each bit of the key can be considered independent to each other when running random instructions in the non-rCore.

·$P$: probability to set the $i^{th}$ bit of the key right.

·$P$: probability to set all bits of the key right simultaneously.

·$\mu_i$: the number of modifications to set the $i^{th}$ bit right.

·$M$: the number of modifications to get the right key.

·$E(\mu_i)$: the average number of $\mu_i$

·$N_i$: the average number of periods to modify the $i^{th}$ bit. (Note that not all the bits can be modified within a single clock period)

·$T_m$: the average number of clock periods to decipher the key by instructions.

$\mu_i$ and $M$ satisfy geometric distribution, respectively. From Probability Theory we will get:

$$E(\mu_i) = \frac{1}{p_i} \quad (1)$$

$$E(M) = \frac{1}{P} = \frac{1}{\Pi_{i=1}^m p_i} = \Pi_{i=1}^m E(\mu_i) \quad (2)$$

$$T_m = \Pi_{i=1}^m E(\mu_i) N_i \geq \left[\min_{1 \leq i \leq m}\{E(\mu_i) N_i\}\right]^m \quad (3)$$

To decipher the traditional logic obfuscation paradigm like logic locking, the hacker needs to use 2^m clock cycles by brute attack to traverse the whole FSM. Unfortunately, the needed clock cycles can be sharply reduced with DIPs found in the process of iteration and the limited key-length due to the constraint of non-negligible overheads. Nevertheless, the get the correct key in our ALL-MASK architecture, the hacker has to spend even more clock cycles by brute attacks due to the following two reasons. One is that the hacker will inevitably re-traverse some states that have been traversed. The other is that most of the bits of the key tend to remain unchanged due to inner mechanism of the non-rCore microarchitecture. Furthermore, as aforementioned, the hacker has to run random IIS to configure all the rGates in rCores right simultaneously since he/she cannot use some pruning algorithms like SAT attack to crack the key into smaller pieces. To say the least, even if the hacker knows the key in some way, it is not a plain task to get the correct keys since the internal nodes are invisible to hackers due to secure-scan interface and the hacker does not have the command of the relationship between the instructions and internal nodes.

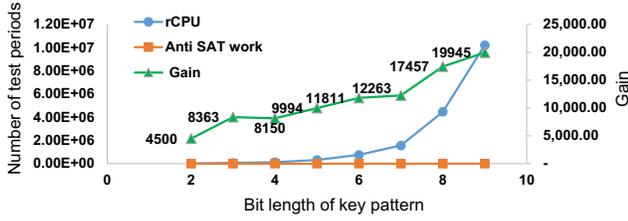

**Fig. 14.** The number of cycles required to traverse the key patterns of anti-SAT-based work [9][10][11] and this work.

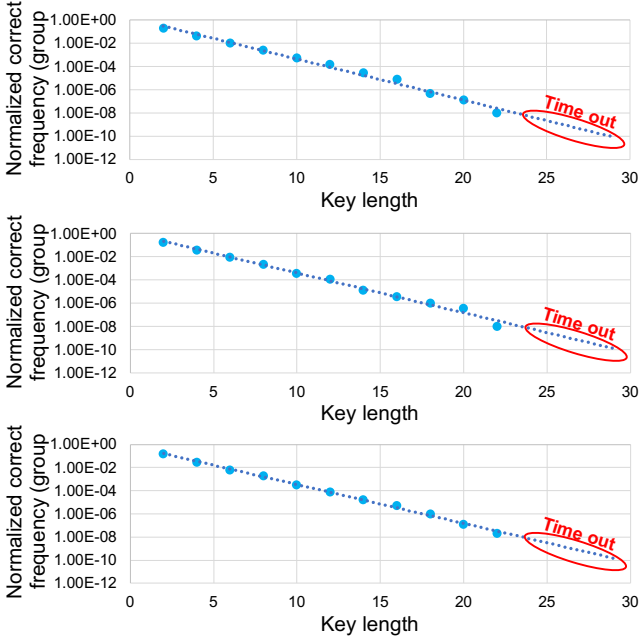

**Fig. 15.** The relationship between the key length and normalized correct frequency in three groups of selected nodes.

This work verifies the model mentioned above by simulation. Experiments are established on a MIPS microprocessor Single-Cycle CPU core to show the relationship between IIS and nodes in the non-rCore. The process of key generation is simulated through the input of random IIS. Fig.14 shows the huge time ratio between traversing the key with IIS and directly traversing through I/O ports as described in previous logic obfuscation paradigm like logic locking. With the key-length increasing, the cost of time for traversing increases exponentially, and the ratio between the traversing time in two different modes increases up to 19,945x for a 9-bit key. Since the key-length could be extend by inserting more rGates due to the negligible overheads, the exponential growth of traversing time can be maintained. Therefore, it is easy for the hacker to breach the endurance limit of FeFETs, leading to the irreversible failure in the ALL-MASK architecture, which enhances the economic expenses of malicious attacks.

To further illustrate the difficulty of constructing the correct key by IIS, we randomly select three groups of nodes in the non-rCore and randomly assume a unique sequence of key. As shown in Fig.15, the results of three different groups (group A, B and C) of selected nodes exhibit similar characteristics. Assume that the length of the key and the number of selected nodes in the non-rCore are both K (usually considerable enough to ensure security, e.g., 64 bits). With the linear increase of the key-length, the simulated cost of time grows exponentially. When the key-length grows longer than 22 bits, we cannot obtain a completely correct key within $10^8$ clock cycles. The experiment also shows that while partial cracking is easy, it is much harder to configure all bits of the key correct. For example, the probability of getting the first four bits correct is approximately 0.04, but the probability of getting the first twenty bits correct drops down exponentially to only $1.3 \times 10^{-7}$ due to the increase of key-length.

Although the locations of the nodes are selected randomly, the results of the three groups are consistent with each other. We can use an exponential curve to predict the growth trend of traversing time with the increase of key-length, as shown in Fig. 14 with the dotted line. Assume that the clock frequency of the non-rCore is 1GHz, as we insert more rGates into rCores and select more nodes in the non-rCore, the corresponding cost of time surges up from about 78 days to about 14,000 years when the key-length rises from 48 to 64. The increase of time becomes an important constraint to prevent key attempts and therefore ensure security. Besides, the endurance of FeFETs further limits multiple trials. As demonstrated above, by combining the traits of instruction level (IIS), microarchitecture level (non-rCore and rCores) and device level (FeFET), ALL-MASK strongly guarantees the security of the multicore system.

### D. Security Analysis and Obfuscation Evaluation

The proposed ALL-MASK method is designed for protecting against IP piracy, together resilient to counterfeiting, reverse engineering, and Trojan. In addition to traditional hardware threats, this section summarizes the resilience to algorithm attack and evaluates the obfuscation effect of the rGates.

*IP piracy and counterfeiting*: As illustrated in section III, ALL-MASK hides the original IP design and extend the functionality at logic gate level. This extension takes the original gate as the base unit and modifies the fixed output into a variable output. Using the configurable-$V_T$ characteristic of FeFETs, ALL-MASK obfuscate the IP with multiple incorrect functions, and the only way to obtain the correct functionality is to get the correct key/IIS. Assuming that a malicious vendor counterfeits a chip with the exact same circuit structure. The most critical part, which is the threshold voltage of each FeFET, is unsure unless the IIS from the non-rCore allows the corresponding FSM to reach the correct state to generate the correct key. Without the correct function of logic gates, the counterfeit chip cannot work well and loses the value.

*Reverse engineering*: Taking the advantage of the TVD characteristic of the FeFETs, ALL-MASK is resilient to the reverse engineering since the structure cannot reveal the obfuscation details. The critical information is hidden in the intrinsic characteristic of the FeFETs and the reverse engineering cannot attack it. Unlike the prior method which may expose the hardware peripheral circuit structure under the attack, the rGates in ALL-MASK are more lightweight and high-density.

*PUF removal attack and Trojan*: PUF is an individual identification that ALL-MASK adopts to control the *Vdd* supply. Each chip is required to pass independent certification to have the access of controlling *Vdd*, which is a necessary condition for changing the ferroelectric polarization state. This uniqueness makes it difficult for hackers to implant hardware trojans, because hackers must be able to modify voltage to modify logic gate functionality. Prior works also propose the protection method against PUF removal attack by using the timing constraint condition [3][48].

Algorithm attack: as analyzed thoroughly in section IV.B and IV.C, ALL-MASK relies on the following two main improvements to enhance the resistance to algorithm attacks. (1) This work makes the key attempts more difficult than ever while attacking, since the key are generated from restricted FSM by inputting IIS. Our analysis in section IV.C shows that it's extremely difficult and it takes significant long time to get the correct key, which is fatal to algorithm attacks through constant pruning key. (2) This work requires the rGates reconfigure simultaneously, which demands the circuit cannot be cracked separately. In Fig. 13(a), the attacker can divide the circuit into three sub-modules and then traverse the possible key separately, which only needs at most $2^1+2^2+2^2 = 10$ cycles to get the correct key. ALL-MASK not only increases the time to obtain the possible key, but also prevent the division and combine the key. As shown in Fig. 13(b), the combinations of key are added to $2^5 = 32$, which is significantly higher for just 5-bit situation. As the key bit is increased, the time of getting one key combination and the number of total key combinations will both grown exponentially.

*Obfuscation effect*: the effect of logic obfuscation is an important part of the evaluating an obfuscation method. Two metrics are adopted in this work: error rate and average hamming distance ratio (HD/Fanout). Error rate represents the error output probability caused by the rGates, and the average HD/Fanout further measures the number of error bits ratio in the outputs. As shown in Fig. 16, this work evaluates nine modules in the ISCAS85 dataset. Since the number of logic gates in modules varies, fewer reconfigurable gates are used for evaluation for small modules and vice versa. The simulation results show that the error rate generally increases as the key length (the replacement rate of the rGate) increases. According to the trend of error rate, it can be predicted that when K rises to a certain value, the error rate will approach and reach 100%. The intuitive interpretation is that the change of the function of the logic gate will lead to the error of the logic output, and the more the change of the logic gate, the greater the possibility of error. The same phenomenon occurs with the average error hamming distance ratio. As the number of logical gate substitutions increases, the number of error bits in the output gradually increases.

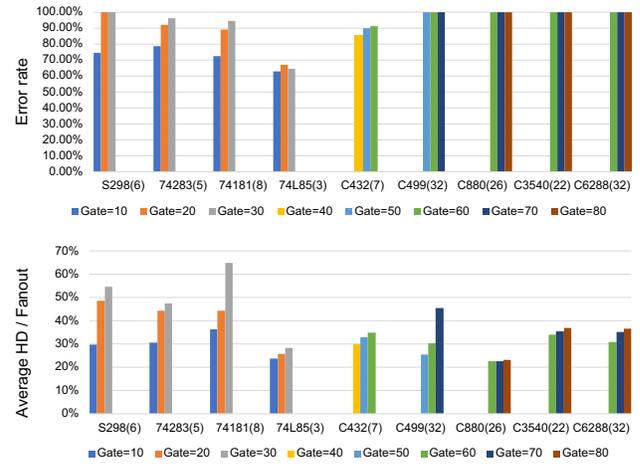

**Fig. 16.** Error rate and average hamming distance for different ISCAS module when K ∈ {10,20, …, 80}.

### E. Overheads

ALL-MASK implements logic obfuscation by using FeFET-based rGates to replace original CMOS logic gates. According to the experimental analysis in the previous section, the effect of logic obfuscation increases exponentially with the number of rGates grows linearly (that is, the linear growth of key-length). However, the increase of the key length is bound to be constrained and challenged by delay, power consumption and area. This section demonstrates to analyze and explain that the overheads of ALL-MASK is significantly lower than that of previous work with the slightly change of the gate structure, which means that there is room to expand the key length to a new level. In this section, the gate-level and module-level overheads are evaluated to demonstrate the outstanding advantages of the FeFET-based obfuscation method in terms of low overheads.

*Gate-level*: in section III.A, ALL-MASK introduces the basic circuit structure of rGate. The two structures make use of the non-volatile characteristic of FeFET so that the FeFETs can remain "always on" state (LVT) or "always off" state (HVT) after the configuration. The two states are similar to "short circuit" and "open circuit", respectively, and the obfuscation logic can be added or deleted to achieve multiple functions. The addition of transistors complicates the structure of the logic gates, and the overheads are the critical evaluation metrics. This work evaluates the latency and energy of the rGates, and the gate-level simulation is done in 65nm CMOS process. We use FeFET models in [49] and compare the CMOS gates with rGates. It should be noted that the two states of FeFETs result in not only different functionality, but also different latency and energy consumption. Therefore, three parts are evaluated for the same number of variables: CMOS gate, LVT rGate and HVT rGate.

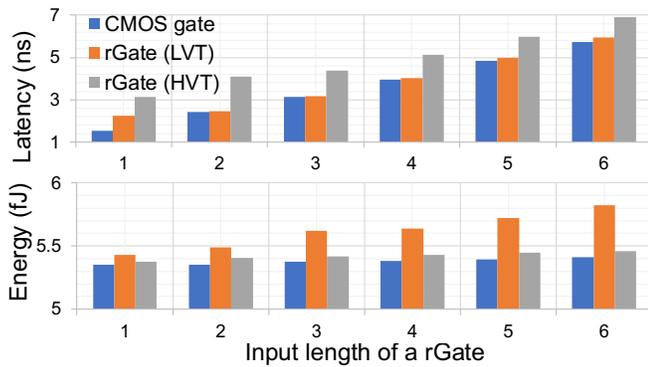

**Fig. 17.** Latency and energy of CMOS logic gate and two function of type-1 rGate (LVT and HVT) for different input variable.

As shown in Fig. 17, the latency increases as the number of variables (that is the number of the transistors) increases, and the latency of HVT rGate is about 1ns higher than LVT rGate. With the input number increases, the latency of LVT rGate is just less than 0.2ns higher than the CMOS gate, which also inspires the designers to make more use of type-A obfuscation (introduced in section III.B) to pursue better timing effects. Similarly, the energy consumption also demonstrates a slightly increase of less than 0.5fJ as the number of variables increases. In this case, the energy of the LVT rGate is higher than the other two situations, and the corresponding type-C obfuscation (introduced in section III.B) is suitable for the energy-harvesting application. The type-2 rGate structure and the corresponding obfuscation characteristic are similar to the previous example, and the design space of the proposed rGates have great potential.

*Module-level*:

As shown in Fig. 13 (c), the replacement strategy of the rGates is related to the critical path. The circuit module has a path with the longest delay, and our replacement principle is to replace the CMOS gate with a reconfigurable gate in the non-critical path, and ensure that the critical path does not change. Another important factor is the key length, in this case the number of rGates, is a very small proportion of the number of CMOS gates in the modules or processors that require encryption. When the proportion of rGates is low, the overheads of power and area brought by these rGates will be weighted and averaged by the overall relatively huge power consumption and area and will be almost negligible. Low proportion does not mean low quantity, and the logic gates after modifying functions can play a critical role in logic obfuscation and cause high error rate and hamming distance.

As discussed in section IV.D, the module logic functions will be greatly changed after the change of logic gate functions, but in fact only a small part of logic gates need to be changed to achieve this goal. In short, not only can a small number of reconfigurable gates lead to a significant increase in obfuscations, but a small percentage of gate costs at the module/processor level are almost insignificant. It is worth noting that there are many circuit modules that need to be encrypted in the actual processor, but this work makes use of the key integration and configuration simultaneously, so that these modules cannot be individually cracked.

## V. DISCUSSION

The position selection strategy of the internal nodes and rGates is a factor to influence the attack resiliency and obfuscation effect. The choice of internal nodes will produce different behaviors for different types of instruction flow due to their different positions, which will affect the whole state transition diagram and key traversal behavior. In this work, three groups of different internal nodes were randomly selected and summarized. Similar exponential prediction properties were found, but there was no guarantee that nodes selected in a specific way or in a more complex scale processor would still show the same properties. As for the rGate position, the function change of logic gate level has a great influence on output error rate and Hamming distance, but how to quantify the degree of this influence is a direction of future research. The significance is that if the designer can get equal or better obfuscation with smaller amount of rGate replacement, the overheads are reduced and the extra key can be used for dummy obfuscation and camouflage, improving security and lowering the cost.

## VI. CONCLUSION

This work has proposed ALL-MASK, a novel logic locking method to enhance the security level of a multicore microarchitecture. We use a core to generate key through a set of specific instructions without the use of extra dedicated key-storage memory module. We consider the complexity and integrity of key acquisition and carry out the corresponding evaluation, proving that it is difficult to obtain the correct key in the actual protection situation of scan-blocked chain, thus improving the security. The key is then used to configure the FeFET-based reconfigurable cores to the target function, and the error rate and Hamming distance are high to enhance the obfuscation effect. Since our design takes advantage of FeFET's low DC power consumption characteristics and high CMOS integration compatibility, the reconfigurable gate is, to the best of our knowledge, the easiest form to retrofit from the original logic gate with minimum overheads. Tests at the module level show that the additional power consumption caused by the logic gate modification is negligible under certain proportional conditions. Future explorations include the key-chosen strategy, IIS optimization, purely-rCore configuration, etc.

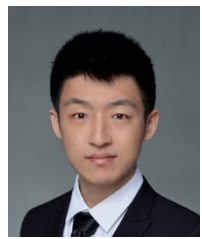

**Jianfeng Wang** (Student Member, IEEE) is currently working toward the B.S. degree in electronic engineering at Tsinghua University, Beijing, China.

His current research interests include emerging circuit design with beyond-CMOS technologies, reconfigurable circuit design and hardware security.

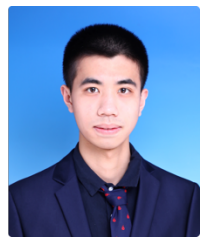

**Zhonghao Chen** (Student Member, IEEE) is currently working toward the B.S. degree in electronic engineering at Tsinghua University, Beijing, China.

His current research interests include memory-centric computing, reconfigurable circuit design and hardware security.

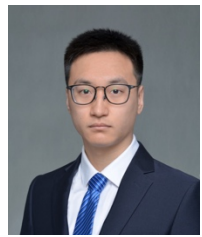

**Jiahao Zhang** (Student Member, IEEE) is currently working toward the B.S. degree in School of Integrated Circuits at Tsinghua University, Beijing, China.

His recent research interests concern chip security and reconfigurable computing chip architectures.

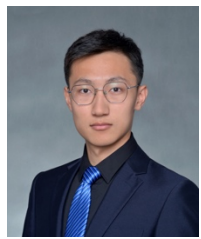

**Enze Ye** (Student Member, IEEE) is currently working toward the B.S. degree in electronic engineering at Tsinghua University, Beijing, China.

His current research interests include reconfigurable circuits and systems.

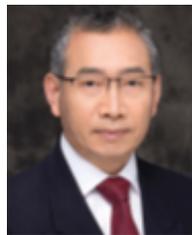

**Huazhong Yang** (Fellow, IEEE) received the B.S. degree in microelectronics and the M.S. and Ph.D. degrees in electronic engineering from Tsinghua University, Beijing, China, in 1989, 1993, and 1998, respectively.

In 1993, he joined the Department of Electronic Engineering, Tsinghua University, where he has been a professor since 1998. He has been in charge of several projects, including projects sponsored by the National Science and Technology Major Project, the 863 Program, NSFC, and several international research cooperations. He has authored or coauthored over 500 technical articles, seven books, and over 180 granted Chinese patents. His research interests include wireless sensor networks, data converters, energy-harvesting circuits, nonvolatile processors, and brain-inspired computing.

Prof. Yang received the Distinguished Young Researcher by NSFC in 2000, the Cheung Kong Scholar by the Chinese Ministry of Education (CME) in 2012, the Science and Technology Award First Prize by the China Highway and Transportation Society in 2016, the Technological Invention Award First Prize by CME in 2019, the Gold Prize of iNEA 2019, and several best paper awards, including ISVLSI 2012, FPGA 2017, NVMSA 2017, and ASP-DAC 2019. He has served as the Chair for the Northern China ACM SIGDA Chapter Science in 2014, the General Co-Chair for ASP-DAC 2020, a Navigating Committee Member for AsianHOST'18, and a TPC Member for ASP-DAC 2005, APCCAS 2006, ICCCAS 2007, ASQED 2009, and ICGCS 2010.

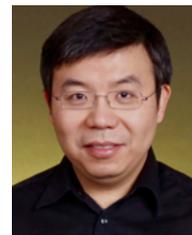

Yongpan Liu (Senior Member, IEEE) received the B.S., M.S., and Ph.D. degrees from the Department of Electronic Engineering, Tsinghua University, Beijing, China, 1999, 2002, and 2007, respectively.

He was a Visiting Scholar with Pennsylvania State University, State College, PA, USA, and the City University of Hong Kong, Hong Kong. He is currently a Professor with the Department of Electronic Engineering, Tsinghua University. He has published over 200 peer-reviewed conference and journal papers and developed several fast sleep/wakeup nonvolatile processors using emerging memory and artificial intelligent accelerators using algorithm-architecture co-optimization. His main research interests include energy-efficient circuits and systems for artificial intelligence, emerging memory devices, and Internet-of-Things (IoT) applications.

Prof. Liu's work has received the Under 40 Young Innovators Award DAC in 2017, the Micro Top Pick in 2016, the Best Paper Award at ASP-DAC 2017 and HPCA 2015, and the Design Contest Awards of ISLPED in 2012 and 2013. He served as the General Chair for AWSSS 2016 and IWCR 2018, the Technical Program Chair for NVMSA 2019, and a Program Committee Member for DAC, DATE, ASP-DAC, ISLPED, ICCD, and A-SSCC. He is an Associate Editor of IEEE Transactions on Computer-Aided Design of

Integrated Circuits and Systems, IEEE Transactions on Circuits and Systems—II: Express Briefs, and IET Cyber-Physical Systems: Theory and Applications.

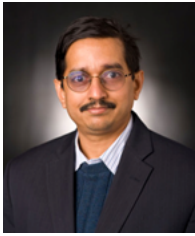

Vijaykrishnan Narayanan (Fellow, IEEE) received the B.S. degree in computer science and engineering from the University of Madras, Chennai, India, in 1993, and the Ph.D. degree in computer science and engineering from the University of South Florida, Tampa, FL, USA, in 1998.

He is currently the Robert Noll Chair Professor of Computer Science and Engineering and Electrical Engineering at Pennsylvania State University, University Park, PA, USA. He is also the Co-Director of the Microsystems Design Laboratory. His current research interests include power-aware and reliable systems, embedded systems, nanoscale devices, and interactions with system architectures, reconfigurable systems, computer architectures, network-on-chips, and domain-specific computing.

Prof. Narayanan was a recipient of several awards, including the University of Madras First Rank in computer science and engineering in 1993, the IEEE Computer Society Richard E. Merwin Award in 1996, the Upsilon Pi Epsilon Award for Academic Excellence in 1997, the Association for Computing Machinery (ACM) Special Interest Group on Design Automation Outstanding New Faculty Award in 2000, the Penn State CSE Faculty Teaching Award in 2002, the IEEE Circuits and Systems (CAS) Society VLSI Transactions Best Paper Award in 2002, and the Penn State Engineering Society Outstanding Research Award in 2006. He was also a recipient of several certificates of appreciation for outstanding service from ACM and the IEEE Computer Society.

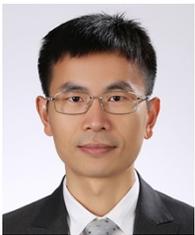

Xueqing Li (Senior Member, IEEE) received the B.S. and Ph.D. degrees from the Department of Electronic Engineering, Tsinghua University, Beijing, China, in 2007 and 2013, respectively.

He is currently an Associate Professor with the Department of Electronic Engineering, Tsinghua University. From 2013 to 2017, he was a Postdoctoral Researcher with the Department of Computer Science and Engineering, Penn State University, University Park, PA, USA. He joined the Department of Electronic Engineering, Tsinghua University, as an Assistant Professor, in 2018. He has more than 100 publications and holds 20 China and U.S. patents. His research interests include low-power circuit design, emerging memory and memory-oriented computing with beyond-CMOS technologies, and high-performance data converter circuit design.

Prof. Li was a recipient of the National Early-Career Award in 2019, the HPCA 2015 Best Paper Award, the 2016 IEEE Micro Top Picks, the ASP-DAC 2017 Best Paper Award, the 2017 IEEE Transactions on Multi-Scale Computing Systems (TMSCS) Best Paper Award, and the 2017 Best Publication Award in SRC/DARPA LEAST Center. He also received several teaching and thesis awards of Tsinghua University and Beijing Municipality. He has served as a TPC or Organizing Committee Member in a few conferences, including ICCAD, ASP-DAC, NVMSA, GLSVLSI, NANOARCH, COINS, CAD-TFT.